\documentclass[showkeys,showpacs,superscriptaddress]{revtex4-2}
\usepackage{amsmath}
\usepackage{amssymb}
\usepackage{graphicx}
\usepackage{float}
\usepackage{xcolor}

\begin{document}

\title{
Linear energy density and the flux of an electric field in Proca tubes
}

\author{
Vladimir Dzhunushaliev
}
\email{v.dzhunushaliev@gmail.com}
\affiliation{
Department of Theoretical and Nuclear Physics,  Al-Farabi Kazakh National University, Almaty 050040, Kazakhstan
}
\affiliation{
Institute of Experimental and Theoretical Physics,  Al-Farabi Kazakh National University, Almaty 050040, Kazakhstan
}
\affiliation{
Academician J.~Jeenbaev Institute of Physics of the NAS of the Kyrgyz Republic, 265 a, Chui Street, Bishkek 720071, Kyrgyzstan
}

\author{Vladimir Folomeev}
\email{vfolomeev@mail.ru}
\affiliation{
Institute of Experimental and Theoretical Physics,  Al-Farabi Kazakh National University, Almaty 050040, Kazakhstan
}
\affiliation{
Academician J.~Jeenbaev Institute of Physics of the NAS of the Kyrgyz Republic, 265 a, Chui Street, Bishkek 720071, Kyrgyzstan
}
\affiliation{
International Laboratory for Theoretical Cosmology, Tomsk State University of Control Systems and Radioelectronics (TUSUR),
Tomsk 634050, Russia
}

\author{
Abylaikhan Tlemisov
}
\email{tlemissov-ozzy@mail.ru}
\affiliation{
Department of Theoretical and Nuclear Physics,  Al-Farabi Kazakh National University, Almaty 050040, Kazakhstan
}
\begin{abstract}
We study cylindrically symmetric solutions within SU(3) non-Abelian Proca theory coupled to a Higgs scalar field.
The solutions describe tubes containing either the flux of a color electric field or the energy flux and momentum.
It is shown that the existence of such tubes depends crucially on the presence of the Higgs field (there are no such solutions without this field).
We examine the dependence of the integral characteristics (linear energy and momentum densities) on the values of the electromagnetic potentials at the center of the tube,
as well as on the values of the coupling constant of the Higgs scalar field. The solutions obtained are topologically trivial and they demonstrate the
dual Meissner effect: the electric field is pushed out by the Higgs scalar field.
\end{abstract}



\date{\today}

\maketitle

\section{Introduction}

In quantum chromodynamics (QCD), it is assumed that
color non-Abelian fields between quark and anti-quark are confined in a tube, due to a strong nonlinear interaction between different components of such fields.
The properties of such a tube are such that outside the tube all fields, and hence the energy density, decrease exponentially with distance.
Inside such a tube, there is a longitudinal electric field connecting quarks and attracting them each other; this is the explanation of quark confinement.
In classical SU(3) non-Abelian Yang-Mills theory uncoupled to another fields, such solutions are apparently absent.
In turn, the lattice calculations in QCD indicate that such configurations of non-Abelian fields do exist.
When other fields are involved, such solutions are already present. For example,
when an electromagnetic field interacts with the Higgs scalar field,
there exist tubes possessing a flux of a magnetic field~-- the well-known solutions found by
Nielsen and Olesen \cite{Nielsen:1973cs}. Non-Abelian flux tube solutions
with the flux of a magnetic field whose force lines  are twisted along the tube axis have been obtained in Ref.~\cite{Brihaye:2007pd}.

Another interesting fact is that such tubes can exist in Proca theories. For example, in  Ref.~\cite{Brihaye:2016pld}, it was shown that there exist gravitating and
nongravitating  Q-tubes supported by a complex vector field with nonlinear terms which can in some sense imitate the self-interaction in non-Abelian Yang-Mills theory.
In Refs.~\cite{Dzhunushaliev:2019sxk,Dzhunushaliev:2020eqa}, the existence of tubes within  SU(3) Proca theory coupled to a Higgs scalar field
has been demonstrated. In those papers,  two types of tube solutions have been found. In the tubes of the first type, there is a flux of a longitudinal color electric field along the tube
created by color charges (quarks) located at  $\pm \infty$. In the tubes of the second type, there is a momentum directed along the tube.
The presence of such a momentum is apparently equivalent to the presence of the energy flux transferred along the tube.

In the present paper we continue investigations in this direction. In doing so,
we calculate such integral characteristics of the solutions like the total flux of the longitudinal electric field,
the linear energy density, and the total momentum passing across the cross section of the tube, depending on the system parameters. Our purpose here is to get configurations from which further insights may be gained about the more complicated non-Abelian gauge theories.
In particular, this applies to the possibility of application of such tubes to explain the phenomenon of confinement in QCD.

\section{Non-Abelian-SU(3)-Proca-Higgs theory}
\label{Proca_Dirac_scalar}

The Lagrangian describing a system consisting of a non-Abelian SU(3) Proca field $A^a_\mu$ interacting with nonlinear scalar field $\phi$ can be taken in the form
(hereafter, we work in units such that $c=\hbar=1$)
\begin{equation}
	\mathcal L =  - \frac{1}{4} F^a_{\mu \nu} F^{a \mu \nu} -
	\frac{\left( \mu^2 \right)^{a b, \mu}_{\phantom{a b,}\nu}}{2}
	A^a_\mu A^{b \nu} +
	\frac{1}{2} \partial_\mu \phi \partial^\mu \phi +
	\frac{\lambda}{2} \phi^2 A^a_\mu A^{a \mu} -
	\frac{\Lambda}{4} \left( \phi^2 - M^2 \right)^2.
\label{2_10}
\end{equation}
Here
$
	F^a_{\mu \nu} = \partial_\mu A^a_\nu - \partial_\nu A^a_\mu +
	g f_{a b c} A^b_\mu A^c_\nu
$ is the field strength tensor for the Proca field, where $f_{a b c}$ are the SU(3) structure constants, $g$ is the coupling constant,
$a,b,c = 1,2, \dots, 8$ are color indices,
$\mu, \nu = 0, 1, 2, 3$ are spacetime indices. The Lagrangian \eqref{2_10} also contains the arbitrary constants $M, \lambda, \Lambda$ and the Proca field mass matrix
$
	\left( \mu^2 \right)^{a b, \mu}_{\phantom{a b,}\nu}
$.

Using \eqref{2_10}, the corresponding field equations can be written in the form
\begin{eqnarray}
	D_\nu F^{a \mu \nu} - \lambda \phi^2 A^{a \mu} &=&
	- \left( \mu^2 \right)^{a b, \mu}_{\phantom{a b,}\nu} A^{b \nu},
\label{2_20}\\
	\Box \phi &=& \lambda A^a_\mu A^{a \mu} \phi +
	\Lambda \phi \left( M^2 - \phi^2 \right) ,
\label{2_30}
\end{eqnarray}
and the energy density is
\begin{equation}
\begin{split}
	\varepsilon = &\frac{1}{2} \left( E^a_i \right)^2 +
	\frac{1}{2} \left( H^a_i \right)^2 -
	\left[
		\left( \mu^2 \right)^{a b, \alpha}_{\phantom{a b,} 0} A^a_\alpha A^b_0 -
		\frac{1}{2} \left( \mu^2 \right)^{a b, \alpha}_{\phantom{a b,} \beta} A^a_\alpha A^{b \beta}
	\right]
	+
	\frac{1}{2} \left( \partial_t \phi \right)^2 +
	\frac{1}{2} \left( \nabla \phi \right)^2
\\
	&
	+\lambda \phi^2 \left[
		\left( A^a_0 \right)^2 - \frac{1}{2} A^a_\alpha A^{a \alpha}
	\right] +
	\frac{\Lambda}{4} \left( \phi^2 - M^2 \right)^2 ,
\label{2_40}
\end{split}
\end{equation}
where $i=1,2,3$ and $E^a_i$ and $H^a_i$ are the components of the electric and magnetic field strengths, respectively.

\section{Proca tube with the flux of the electric field}
\label{electric_flux_section}

To obtain a tube filled with a longitudinal color electric field, we choose the {\it Ans\"{a}tze}~\cite{Obukhov:1996ry, Singleton:1995xc}
\begin{equation}
	A^2_t = \frac{h(\rho)}{g} , \;
	A^5_z = \frac{v(\rho)}{g} , \;
	A^7_\varphi = \frac{\rho w(\rho)}{g} , \;
 	\phi = \phi(\rho),
\label{3_a_10}
\end{equation}
where $\rho, z,$ and $\varphi$ are cylindrical coordinates. In Refs.~\cite{Dzhunushaliev:2019sxk,Dzhunushaliev:2020eqa}, it was shown that solutions describing
such configurations do exist. Here, we would like to study in more detail the dependence of the flux of the electric field along the tube on the parameters determining the solutions.

To simplify the problem, we will consider field configurations with a zero potential $A^7_\varphi = 0$.
In this case we have the following nonzero components of the electric and magnetic field intensities:
\begin{equation}
	E^2_\rho = - \frac{h^\prime}{g} , \quad
	E^7_z = \frac{h v}{2 g} , \quad
	H^5_\varphi = - \frac{\rho v^\prime}{g}.
\label{3_a_30}
\end{equation}
For such a tube, the energy density \eqref{2_40} yields
\begin{equation}
\begin{split}
	g^2 \varepsilon =  \frac{\left( h^\prime \right)^2}{2} +
	\frac{\left( v^\prime \right)^2}{2} +
	g^2 \frac{\left( \phi^\prime \right)^2}{2} +
	\frac{h^2 v^2}{8} -
	\frac{\mu_1^2}{2} h^2 - \frac{\mu_2^2}{2} v^2
	+ \frac{\lambda}{2} \phi^2 \left(
		h^2 + v^2
	\right) +
			\frac{g^2 \Lambda}{4} \left( \phi^2 - M^2 \right)^2
\end{split}
\label{3_a_50}
\end{equation}
with the following components of the Proca field mass matrix: $\mu_{1}^2=\left( \mu^2 \right)^{2 2, t}_{\phantom{a b,}t}$ and
 $\mu_{2}^2=\left( \mu^2 \right)^{5 5, z}_{\phantom{a b,}z}$.

Substituting the potentials \eqref{3_a_10} in Eqs.~\eqref{2_20} and \eqref{2_30} and introducing the dimensionless variables
 $\tilde \phi = \phi \sqrt{\lambda} / \phi_0$,
$\tilde h = h / \phi_0$,
$\tilde v = v / \phi_0$,
$\tilde M = M \sqrt{\lambda} / \phi_0$,
$\tilde \lambda = \lambda / g^2$,
$\tilde \Lambda = \Lambda / \lambda$,
$\tilde \mu_{1,2} = \mu_{1,2} / \phi_0$, and
$x = \rho \phi_0$ [here $\phi_0$ is the central value of the scalar field], we get the following set of equations:
\begin{eqnarray}
  \tilde h'' + \frac{\tilde h'}{x} &=& h
  \left(
  	\frac{{\tilde v}^2}{4} + {\tilde \phi}^2 -
  	\tilde \mu_1^2
  \right) ,
\label{3_a_93}\\
  \tilde v'' + \frac{\tilde v'}{x} &=& v
  \left(
	  - \frac{\tilde h^2}{4} + {\tilde \phi}^2 -
	  \tilde \mu_2^2
  \right) ,
\label{3_a_96}\\
  \tilde \phi'' + \frac{\tilde \phi'}{x} &=&
  \tilde \phi
  \left[
  	\tilde \lambda \left( - \tilde h^2 + \tilde v^2 \right) +
  	\tilde \Lambda \left( {\tilde \phi}^2 - {\tilde M}^2 \right)
  \right].
\label{3_a_99}
\end{eqnarray}
Here,  the prime denotes differentiation with respect to the dimensionless radius $x$.
We seek a solution to Eqs.~\eqref{3_a_93}-\eqref{3_a_99} in the vicinity of the origin of coordinates in the form
\begin{eqnarray}
	\tilde h(x) &=& \tilde h_0 + \tilde h_2 \frac{x^2}{2} + \dots \quad \text{with} \quad
\tilde h_2 = \frac{\tilde h_0}{2} \left(
		\frac{\tilde v_0^2}{4} + \tilde \phi_0^2 - \tilde \mu^2_1
	\right) ,
\label{3_a_100}\\	
	\tilde v(x) &=& \tilde v_0 + \tilde v_2 \frac{x^2}{2} + \dots \quad \text{with} \quad
\tilde v_2 = \frac{\tilde v_0}{2} \left(
		- \frac{\tilde h_0^2}{4} + \tilde \phi_0^2 - \tilde \mu^2_2
	\right) ,
\label{3_a_110}\\
	\tilde \phi(x) &=& \tilde \phi_0 + \tilde \phi_2 \frac{x^2}{2} + \dots \quad \text{with} \quad
\tilde \phi_2 = \frac{\tilde \phi_0}{2} \left[
		\tilde \lambda \left( - \tilde h_0^2 + \tilde v_0^2 \right) +
		\tilde \Lambda \left(
			{\tilde \phi}_0^2 - {\tilde M}^2
	\right)
	\right],
\label{3_a_120}
\end{eqnarray}
where the expansion coefficients $\tilde h_0, \tilde v_0$, and $\tilde \phi_0$  are arbitrary.

The asymptotic behavior of the functions $\tilde h, \tilde v$, and $\tilde \phi$, which follows from Eqs.~\eqref{3_a_93}-\eqref{3_a_99}, is
\begin{equation}
	\tilde h(x) \approx \tilde h_{\infty}
	\frac{e^{- x \sqrt{\tilde M^2 - \tilde \mu^2_1}}}{\sqrt x},\,\,
	\tilde v(\rho) \approx \tilde v_{\infty}
	\frac{e^{- x \sqrt{\tilde M^2 - \tilde \mu^2_2}}}{\sqrt x},\,\,
	\tilde \phi \approx \tilde M - \tilde \phi_\infty
	\frac{e^{- x \sqrt{2 \tilde \Lambda \tilde M^2}}}{\sqrt x} ,
\label{3_a_160}
\nonumber
\end{equation}
where $\tilde h_{\infty}, \tilde v_{\infty}$, and $\tilde \phi_\infty$ are integration constants.

\begin{figure}[H]
\centering
	\includegraphics[width=1\linewidth]{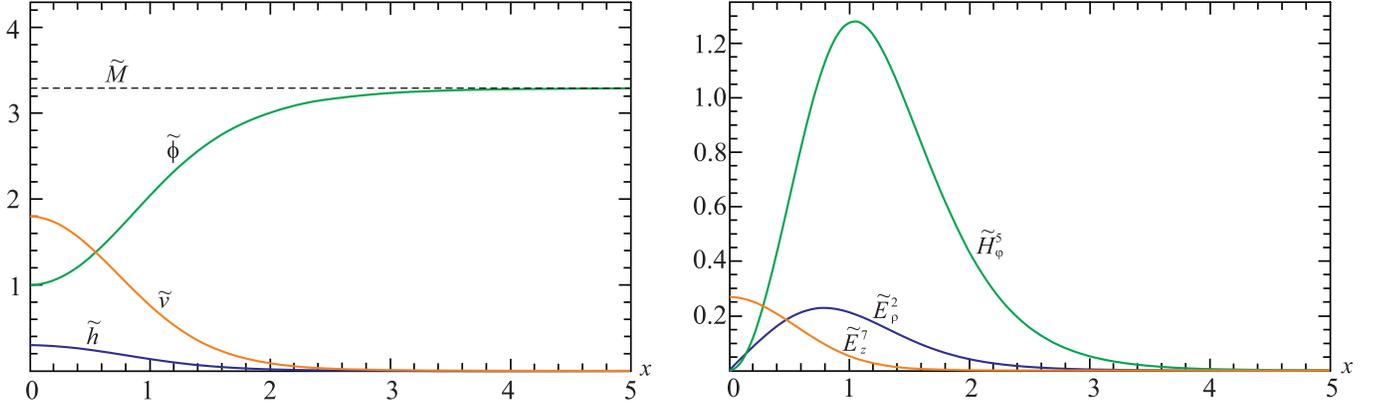}
\caption{The typical profiles of the functions $\tilde h(x), \tilde f(x)$, and $\tilde \phi(x)$ and
of the dimensionless fields $\tilde E^2_\rho$,
$\tilde E^7_z$, and $\tilde H^5_\varphi$
for $\tilde M = 3.29316$,
$\tilde \mu_1 =2.19828$, $\tilde \mu_2 =2.10505$,
$\tilde \phi_0=1.0, \tilde h_0=0.3, \tilde v_0 = 1.8, \tilde \lambda =2.0, \tilde \Lambda =0.1$.
}
\label{E_flux_h_v_phi_E_H_fields}
\end{figure}

The derivation of solutions to the set of equations~\eqref{3_a_93}-\eqref{3_a_99} is an eigenvalue problem for the parameters $\tilde \mu_{1}, \tilde \mu_{2}$, and $\tilde M$.
The numerical solution describing the behavior of the Proca field potentials and of the corresponding electric and magnetic fields is given in Fig.~\ref{E_flux_h_v_phi_E_H_fields}.
The behavior of the electric field $E^7_z$ indicates that we are dealing with the dual Meissner effect:
this field is pushed out by the scalar field $\phi$. Notice also that the tube solutions obtained are topologically trivial, in contrast to the Nielsen-Olesen solution.

\begin{figure}[H]
\centering
	\includegraphics[width=1\linewidth]{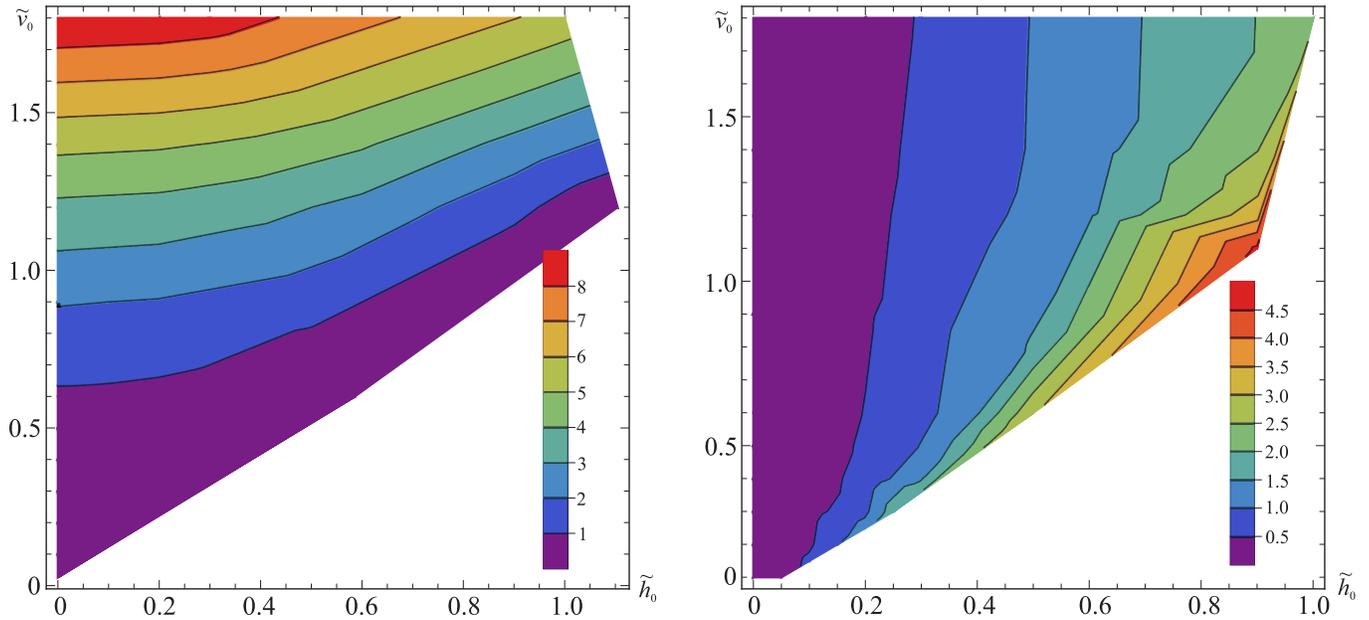}
\caption{The contour profiles of the dimensionless linear energy density
$\tilde{\mathcal E } (x)$ (left panel) and of the dimensionless flux of the longitudinal electric field $\tilde \Phi_z (x)$ (right panel)
as functions of the parameters $\tilde h_0$ and $\tilde v_0$
for the tube with the flux of the longitudinal electric field.}
\label{lin_energy_dens_electric_FT_electric_flux}
\end{figure}

Let us define the linear energy density  $\mathcal E $ and the total flux $\Phi_z$ of the longitudinal electric field $E^7_z$
transferred across the cross section of the flux tube as follows:
\begin{align}
	\mathcal E \equiv \frac{\phi_0^2}{g^2} \tilde{\mathcal E} & =
	2 \pi \frac{\phi_0^2}{g^2}
	\int \limits_0^\infty x \tilde \epsilon d x =
	2 \pi \left(
		\frac{\phi_0/  \Lambda_{\text{QCD}}}{{g^\prime}}
	\right)^2 \hbar c \Lambda^2_{\text{QCD}}
	\int \limits_0^\infty x \tilde \epsilon d x ,
\label{3_a_170}\\
	\Phi_z \equiv \frac{\phi_0^2}{g^2} \tilde \Phi_z & =
	2 \pi \frac{\phi_0^2}{g^2}
	\int \limits_0^\infty x \tilde E^7_z d x =
	2 \pi  \left(
		\frac{\phi_0/  \Lambda_{\text{QCD}}}{{g^\prime}}
		\right)^2 \hbar c \Lambda^2_{\text{QCD}}
	\int \limits_0^\infty x \tilde E^7_z d x,
\label{3_a_180}
\end{align}
where ${g'}^2=g^2 \hbar c$ is the dimensionless coupling constant, the tilde sign denotes that the corresponding quantities are dimensionless, and $\Lambda_{\text{QCD}}$ is a characteristic parameter coming from QCD. The integral characteristics of these quantities are shown in Fig.~\ref{lin_energy_dens_electric_FT_electric_flux}.
The analysis of the results shown in Fig.~\ref{lin_energy_dens_electric_FT_electric_flux}
permits us to assume that
\begin{itemize}
\item when $v_0 \rightarrow 0, h_0 = \text{const}$ the linear energy and momentum densities
$\tilde{\mathcal E } (x), \tilde \Phi_z (x) \rightarrow 0$;
\item when $v_0 \rightarrow \infty, h_0 = \text{const}$ the linear energy and momentum densities
$\tilde{\mathcal E } (x), \tilde \Phi_z (x) \rightarrow \infty$;
\item when $h_0 \rightarrow 0, v_0 = \text{const}$ the linear energy and momentum densities
$\tilde{\mathcal E } (x), \tilde \Phi_z (x) \rightarrow \text{const}$;
\item when $h_0 \rightarrow \infty, v_0 = \text{const}$ the linear energy and momentum densities
$\tilde{\mathcal E } (x), \tilde \Phi_z (x) \rightarrow 0$.
\end{itemize}

Note that, unfortunately, the technical difficulties of numerical solving the set of equations~\eqref{3_a_93}-\eqref{3_a_99}
do not permit one to study the dependencies
$\tilde{\mathcal E } (x)$ and $\tilde \Phi_z (x)$ on the parameters $\tilde h_0$ and $\tilde v_0$ in more detail. The reason is that for small values of
$\tilde v_0$  the calculation accuracy implemented in Wolfram Mathematica does not permit one to find the eigenvalues $\tilde \mu_{1,2}$ and $\tilde M$.

\section{Proca tubes with the momentum density}
\label{momentum_flux}

In the previous section, we have considered tubes  filled with a longitudinal electric field. Such a tube is described by a non-Abelian Proca field
sourced by quarks located at $\pm \infty$. In this section, we consider a tube containing nonzero flux of the Poynting vector, whose presence results in the fact that
there is the energy flux, and hence the momentum directed from one source (located at $-\infty$) to another one (located at $+\infty$).

For this case, we choose the {\it Ans\"{a}tze}
\begin{equation}
	A^5_t = \frac{f(\rho)}{g} , \;
	A^5_z = \frac{v(\rho)}{g} , \;
	A^7_\varphi = \frac{\rho w(\rho)}{g} , \;
 	\phi = \phi(\rho),
\label{3_b_10}
\end{equation}
 which give the following components of the electric and magnetic field intensities:
\begin{eqnarray}
	E^2_\varphi &=& \frac{\rho f w}{2 g}, \quad
	E^5_\rho = - \frac{f^\prime}{g} , \quad
\label{3_b_20}\\
	H^2_\rho &=& - \frac{v w}{2 g}, \quad
	H^5_\varphi = - \frac{\rho v^\prime}{g}, \quad
	H^7_z = \frac{1}{g} \left(
		w^\prime + \frac{w}{\rho}
	\right) .
\label{3_b_30}
\end{eqnarray}
In this case the Poynting vector
\begin{equation}
	S^i = \frac{\epsilon^{i j k}}{\sqrt{\gamma}} E^a_j H^a_k
\label{3_b_35}
\end{equation}
is already nonzero [see Eqs.~\eqref{4_a_60} and \eqref{3_b_50} below].

Substituting the potentials \eqref{3_b_10} in Eqs.~\eqref{2_20} and \eqref{2_30} and using the dimensionless variables
 given before Eq.~\eqref{3_a_93}, we derive the following equations:
\begin{eqnarray}
  \tilde f'' + \frac{\tilde f'}{x} &=& \tilde f
  \left(
  	\frac{\tilde w^2}{4} + {\tilde \phi}^2 - \tilde \mu_1^2
  \right) ,
\label{3_b_100}\\
  \tilde v'' + \frac{\tilde v'}{x} &=& \tilde v
  \left(
	  \frac{\tilde w^2}{4} + {\tilde \phi}^2 - \tilde \mu_2^2
  \right) ,
\label{3_b_110}\\
  \tilde w'' + \frac{\tilde w'}{x} - \frac{\tilde w}{x^2} &=&
  \tilde w \left(
  	- \frac{\tilde f^2}{4} + \frac{\tilde v^2}{4} + {\tilde \phi}^2 -
  	\tilde \mu_3^2
  \right),
\label{3_b_120}\\
  \tilde\phi'' + \frac{\tilde \phi'}{x} &=&
  \tilde \phi
  \left[
  	\tilde \lambda \left( - \tilde f^2 + \tilde v^2 + \tilde w^2 \right) +
  	\tilde \Lambda \left( {\tilde \phi}^2 - {\tilde M}^2 \right)
  \right],
\label{3_b_130}
\end{eqnarray}
where $\tilde f = f / \phi_0$,  $\tilde w = w / \phi_0$, and we have introduced the component of the Proca field mass matrix $\mu_{3}^2=\left( \mu^2 \right)^{7 7, \varphi}_{\phantom{a b,}\varphi}$. This set of equations has a cylindrically symmetric solution describing a tube with nonzero momentum density and energy flux (the Poynting vector).

Our purpose is to study the dependence of the linear momentum density
$
	\int \vec S d \vec \sigma
$ on the boundary conditions $\tilde f_0, \tilde v_0, \tilde w_0$ given at the center and the parameter $\tilde\Lambda$.
Unfortunately, the number of these parameters is too large to investigate this dependence in detail;
our goal will therefore be to reduce this number. To do this, in the next subsection, we examine the case with $\tilde w = 0$.
Such a restriction results in the fact that we will actually deal with Proca electrodynamics, i.e., with $U(1)$ massive electrodynamics possessing
$U(1)$ group spanned on the $\lambda_5$ Gell-Mann matrix.

\subsection{Abelian Proca tubes}
\label{Abelain_Proca_FT}

Consider the simplest case when the component of the potential $A^7_\varphi = 0$.
In this case we deal with massive (Proca) electrodynamics.

Let us examine the simplest particular case where
$\tilde f = h \sinh \xi, \tilde v = h \cosh \xi$ with $\xi = \text{const.}$ and
$\tilde \mu_1 = \tilde \mu_2 = \tilde \mu$. In this case the set of equations
\eqref{3_b_100}-\eqref{3_b_130} is split as follows. Eq.~\eqref{3_b_100} takes the form of the Schr\"{o}dinger equation,
\begin{equation}
  - h^{\prime \prime} - \frac{h^\prime}{x} +
  h {\tilde \phi}^2 	= \tilde \mu^2 h,
\label{4_a_10}
\end{equation}
where the function ${\tilde \phi}^2$ plays the role of
the effective potential for the ``wave function'' $h$.
In order to ensure a regular solution of this equation, it is necessary that the effective potential would possess a well. In this case Eq.~\eqref{4_a_10} must be solved as an eigenvalue problem for the parameter $\tilde \mu^2$ with the eigenfunction $h$.

The remaining equation \eqref{3_b_130} for the function $\tilde \phi$ is  then
\begin{equation}
  \tilde \phi'' + \frac{\tilde \phi'}{x} = \tilde \phi
  \left[
  	\tilde \lambda h^2 +
  	\tilde \Lambda \left( {\tilde \phi}^2 - {\tilde M}^2 \right)
  \right] .
\label{4_a_20}
\end{equation}
Introducing new variables $\bar x = x \tilde \lambda$ and
$\bar \phi = \tilde \phi / \sqrt{\tilde \lambda}$ and the constants
$
	\bar \mu = \tilde \mu / \sqrt{\tilde \lambda} ,
	\bar M = \tilde M  / \sqrt{\tilde \lambda}
$,
Eqs.~\eqref{4_a_10} and \eqref{4_a_20} can be recast in the form
\begin{align}
	 - h^{\prime \prime} - \frac{h^\prime}{\bar x} +
	   {\bar \phi}^2 h	= & \bar \mu^2 h,
\label{4_a_23}\\
  \bar \phi'' + \frac{\bar \phi'}{\bar x} = & \bar \phi
  \left[
  	h^2 + \tilde \Lambda \left( {\bar \phi}^2 - {\bar M}^2 \right)
  \right],
\label{4_a_26}
\end{align}
where the prime denotes differentiation with respect to $\bar x$.

We will seek regular solutions possessing a finite linear energy density. This means that asymptotically (as $\bar x \rightarrow \infty$) the function
$
	h(\bar x) \rightarrow 0
$.
Then, taking into account the positiveness of the effective potential
$ {\bar \phi}^2 $, one can conclude that the function
$\bar \phi$ must go to a constant, and Eq.~\eqref{4_a_26} implies that this constant is $\bar M$, i.e.,
$
	\bar \phi \rightarrow \bar M	\text{ as } \bar x \rightarrow \infty
$.

We seek solutions of Eqs.~\eqref{4_a_23} and \eqref{4_a_26} in the vicinity of the origin of coordinates in the form
\begin{align}
	h(\bar x) =& h_0 + h_2 \frac{\bar x^2}{2} + \dots \quad \text{with} \quad
	h_2 = \frac{h_0}{2} \left(
		 \bar \phi_0^2 - \bar \mu^2
	\right) ,
\label{4_a_30}\\	
	\bar \phi(\bar x) =& \bar \phi_0 +
	\bar \phi_2 \frac{\bar x^2}{2} + \dots \quad \text{with} \quad
\bar \phi_2 = \frac{\bar \phi_0}{2} \left[h_0^2+\tilde \Lambda \left(
		{\bar \phi}_0^2 - {\bar M}^2 \right)\right],
\label{4_a_40}
\end{align}
where the expansion coefficients $h_0$ and $\bar \phi_0$ are arbitrary.

In turn, the asymptotic behavior of the functions is
$$
	h(\bar x) \approx h_{\infty}
	\frac{e^{- \bar x \sqrt{\bar M^2 - \bar \mu^2}}}{\sqrt{\bar x}},\quad
	\bar \phi(\bar x) \approx \bar M - \bar \phi_\infty
	\frac{e^{- \bar x \sqrt{2 \tilde \Lambda \bar M^2}}}{\sqrt{\bar x}} ,
$$
where $h_{\infty}$ and $\bar \phi_\infty$ are integration constants.

\begin{figure}[H]
\centering
	\includegraphics[width=1\linewidth]{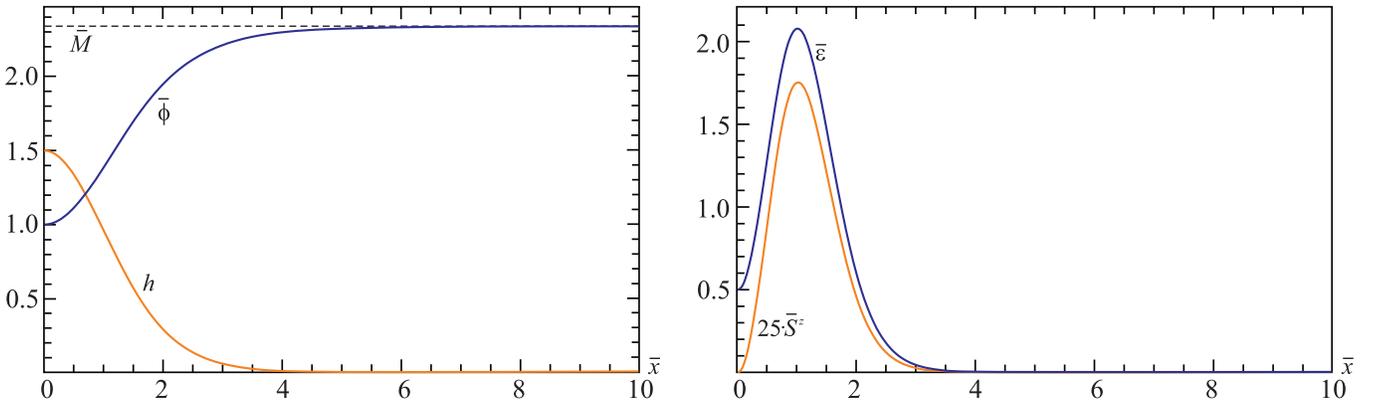}
\caption{The typical profiles of the functions
 $h(\bar x)$ and $\bar\phi(\bar x)$
 and of the dimensionless energy density
$\bar \varepsilon(\bar x)$ and dimensionless momentum density
$\bar S^z (\bar x)$
 for $\bar M = 2.33824$,
$\bar \mu =1.67446587$, $\bar \phi_0=1.0, h_0=1.5, \tilde \Lambda =0.1$, $\xi=0.1$.
}
\label{h_phi_x_energy_density_Abelian}
\end{figure}

The typical profiles of the potential $h(\bar x)$ and of the scalar $\bar\phi(\bar x)$, obtained numerically, are shown in Fig.~\ref{h_phi_x_energy_density_Abelian}. Also, this figure shows
the corresponding graph for the dimensionless energy density
\begin{equation}
\begin{split}
	\bar \varepsilon \equiv \frac{g^6}{\lambda^2\phi^4_0} \varepsilon =&
	\frac{\sinh^2 \xi + \cosh^2 \xi}{2} \left( h^\prime \right)^2 +
	\frac{\left( \bar{\phi}^\prime \right)^2}{2} -
	\frac{
		\bar \mu^2 \left(
			\sinh^2 \xi + \cosh^2 \xi
		\right)
	}{2} h^2 + \frac{
		\bar \mu^2 \left(
			\sinh^2 \xi + \cosh^2 \xi
		\right) 	
	}{2} {\bar \phi}^2 h^2 +
\\
	&
	\frac{\tilde \Lambda}{4}
	\left( \bar \phi^2 - \bar M^2 \right)^2.
\label{4_a_50}
\end{split}
\end{equation}

Substitution of the components of electric and magnetic fields \eqref{3_b_20} and \eqref{3_b_30} in Eq.~\eqref{3_b_35} yields the following expression for the Poynting vector:
\begin{equation}
	S^z = \frac{1}{g^2}
	\frac{d f}{d \rho} \frac{d v}{d \rho}  =
	\frac{\lambda^2\phi^4_0}{g^6} \sinh \xi \cosh \xi
	\left( h^\prime \right)^2 =
	\frac{\lambda^2\phi^4_0}{g^6} \bar S^z=
	\lambda^2\left(\frac{\phi_0^4/\Lambda_{\text{QCD}}^4}{g'^6}\right)\hbar^3c^3\Lambda_{\text{QCD}}^4\bar S^z ,
\label{4_a_60}
\end{equation}
whose typical behavior is given in Fig.~\ref{h_phi_x_energy_density_Abelian}.
Note that the presence of the gradient terms $f^\prime$ and $v^\prime$ means that (as already pointed out above) we are dealing with massive Proca electrodynamics.

\begin{figure}[H]
\centering
	\includegraphics[width=1\linewidth]{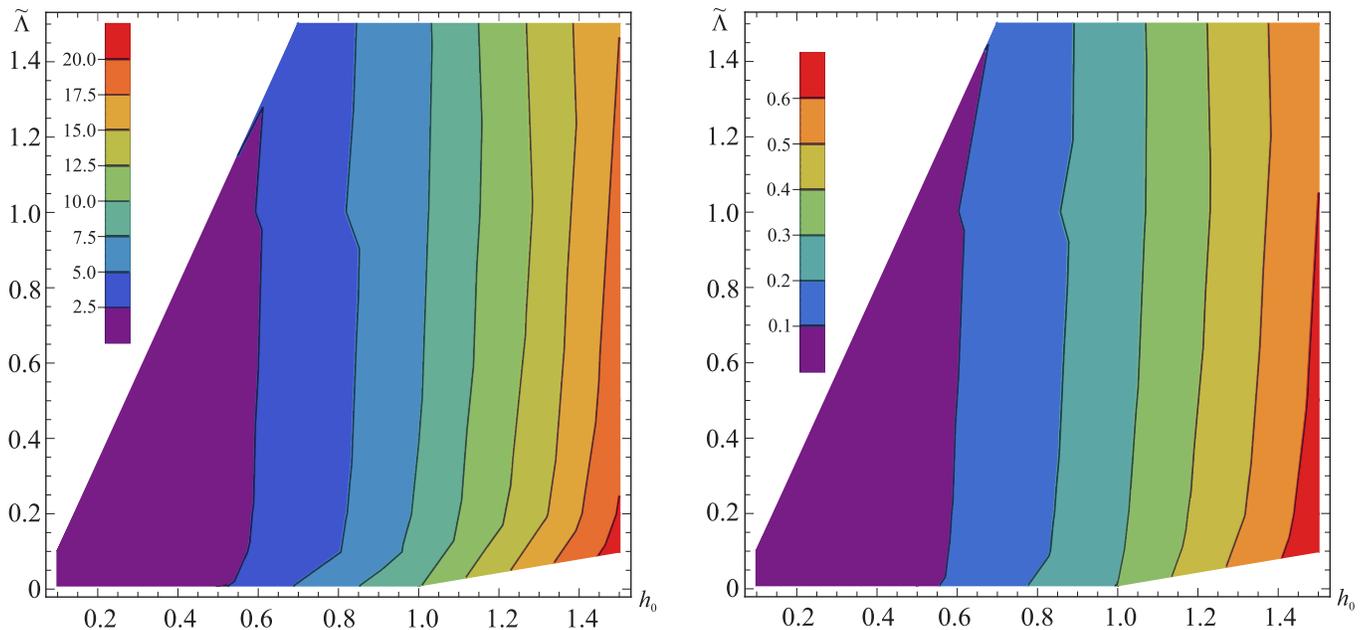}
\caption{The contour profiles of the dependence of the linear energy density $\bar{\mathcal E}$ (left panel)
and of the linear momentum density $\bar \Pi^z$ (right panel) on the parameters  $h_0$ and $\tilde \Lambda$.
}
\label{total_energy_momentum_FT_momentum}
\end{figure}

In our study, the integral characteristics are of greatest interest. Namely, these are the linear energy density
\begin{equation}
		\bar{\mathcal E} = 	\frac{g^2}{\phi_0^2} \mathcal E = 2 \pi
	\int \limits_0^\infty 	\bar{x}  \bar{\varepsilon} d \bar{x}
\label{4_a_70}
\end{equation}
and the linear momentum density across the cross section of the tube
\begin{equation}
	\bar \Pi^z =  	\frac{g^2}{\phi_0^2} M^z = 2 \pi
		\int \limits_0^\infty 	\bar{x}  \bar{S^z} d \bar{x}
\label{4_a_80}
\end{equation}
shown in Fig.~\ref{total_energy_momentum_FT_momentum} as functions of the parameters $h_0$ and $\tilde \Lambda$.
The numerical calculations indicate that both these integral characteristics depend weakly on the coupling constant $\tilde \Lambda$.

Thus, we have demonstrated in this subsection that in Proca electrodynamics there can exist tubes
filled with constant electric and magnetic fields creating the energy flux (or, equivalently, the momentum) directed along the tube axis.
This situation differs in principle from what happens  in an electromagnetic wave where the energy flux (and the momentum)
is created by time-dependent electric and magnetic fields.
The existence of such a tube possessing the energy flux and momentum depends in principle on the presence of the Higgs scalar field.

Based on the analysis of numerical solutions of the corresponding equations, we can suppose that
\begin{itemize}
\item the linear energy, $\mathcal E$, and momentum, $\Pi^z$, densities depend weakly on the coupling constant of the Higgs scalar field $\Lambda$;
\item when $h_0 \rightarrow 0$ the quantities $\mathcal E, \Pi^z \rightarrow 0$;
\item when $h_0 \rightarrow \infty$ the quantities $\mathcal E, \Pi^z \rightarrow \infty$.
\end{itemize}

\subsection{Non-Abelian Proca tubes}

Consider the case of
$\tilde f = \tilde v, \tilde w \neq 0$ and
$\tilde \mu_1 = \tilde \mu_2 = \tilde \mu$. In this case the set of equations
\eqref{3_b_100}-\eqref{3_b_130} is split as follows. Eq.~\eqref{3_b_100} takes the form of the Schr\"{o}dinger equation,
\begin{equation}
  - \tilde f'' - \frac{\tilde f'}{x} + \tilde f U_{\tilde f, \text{eff}}	=
  \tilde \mu^2 \tilde f,
\label{4_b_10}
\end{equation}
where the effective potential for the ``wave function'' $\tilde f$ is
\begin{equation}
	U_{\tilde f, \text{eff}} = \frac{\tilde w^2}{4} + {\tilde \phi}^2 .
\label{4_b_20}
\end{equation}
As in the previous subsection, Eq.~\eqref{4_b_10} must be solved as an eigenvalue problem for the parameter $\tilde \mu^2$ with the eigenfunction $\tilde f$.

The remaining equations \eqref{3_b_120} and \eqref{3_b_130} for the functions
$\tilde w$ and $\tilde \phi$ are then
\begin{eqnarray}
  - \tilde w'' - \frac{\tilde w'}{x} + \tilde w U_{\tilde w, \text{eff}} &=& \tilde \mu_3^2 \tilde w ,
\label{4_b_30}\\
  \tilde \phi'' + \frac{\tilde \phi'}{x} &=&
  \tilde \phi
  \left[
  	\tilde \lambda \tilde w^2 +
  	\tilde \Lambda \left( {\tilde \phi}^2 - {\tilde M}^2 \right)
  \right],
\label{4_b_40}
\end{eqnarray}
and they do not already contain the function $\tilde f$. The effective potential, which appears in Eq.~\eqref{4_b_30}, is
\begin{equation}
	U_{\tilde w, \text{eff}} = \frac{1}{x^2} + {\tilde \phi}^2.
\label{4_b_50}
\end{equation}
The equation~\eqref{4_b_30} has the form of the Schr\"{o}dinger equation with the ``wave function'' $\tilde w$ and with the ``energy'' $\tilde \mu_3^2$. This means that it will have a regular solution only if the effective potential $U_{\tilde w, \text{eff}}$ possesses a well.

We seek solutions of Eqs.~\eqref{4_b_10}, \eqref{4_b_30}, and \eqref{4_b_40} in the vicinity of the origin of coordinates in the form
\begin{eqnarray}
	\tilde f(x) &=& \tilde f_0 + \tilde f_2 \frac{x^2}{2} + \dots \quad \text{with} \quad \tilde f_2 = \frac{\tilde f_0}{2} \left(
		 \tilde \phi_0^2 - \tilde \mu^2
	\right) ,
\label{4_b_60}\\	
	\tilde w(x) &=& \tilde w_1 x + \tilde w_3 \frac{x^3}{3 !} + \dots \quad \text{with} \quad \tilde w_3 = \frac{2}{3}\tilde w_1 \left(\tilde \phi_0^2 -\tilde \mu_3^2 \right),
\label{4_b_70}\\
	\tilde \phi(x) &=& \tilde \phi_0 + \tilde \phi_2 \frac{x^2}{2} + \dots \quad \text{with} \quad
\tilde \phi_2 = \frac{\tilde \phi_0}{2} \tilde \Lambda \left(
		{\tilde \phi}_0^2 - {\tilde M}^2 \right),
\label{4_b_80}
\end{eqnarray}
where the expansion coefficients $\tilde f_0, \tilde \phi_0$, and $\tilde w_1$ are arbitrary.

In turn, the asymptotic behavior of the functions is
$$
	\tilde f(x) = \tilde v(x) \approx \tilde f_{\infty}
	\frac{e^{- x \sqrt{\tilde M^2 - \tilde \mu^2}}}{\sqrt x},\quad
	\tilde w(x) \approx \tilde w_{\infty}
	\frac{e^{- x \sqrt{\tilde M^2 - \tilde \mu^2_3}}}{\sqrt x},\quad
	\tilde \phi \approx \tilde M - \tilde \phi_\infty
	\frac{e^{- x \sqrt{2 \tilde \Lambda \tilde M^2}}}{\sqrt x} ,
$$
where $\tilde f_{\infty}, \tilde w_{\infty}$, and $\tilde \phi_\infty$ are integration constants.

\begin{figure}[H]
\centering
	\includegraphics[width=1\linewidth]{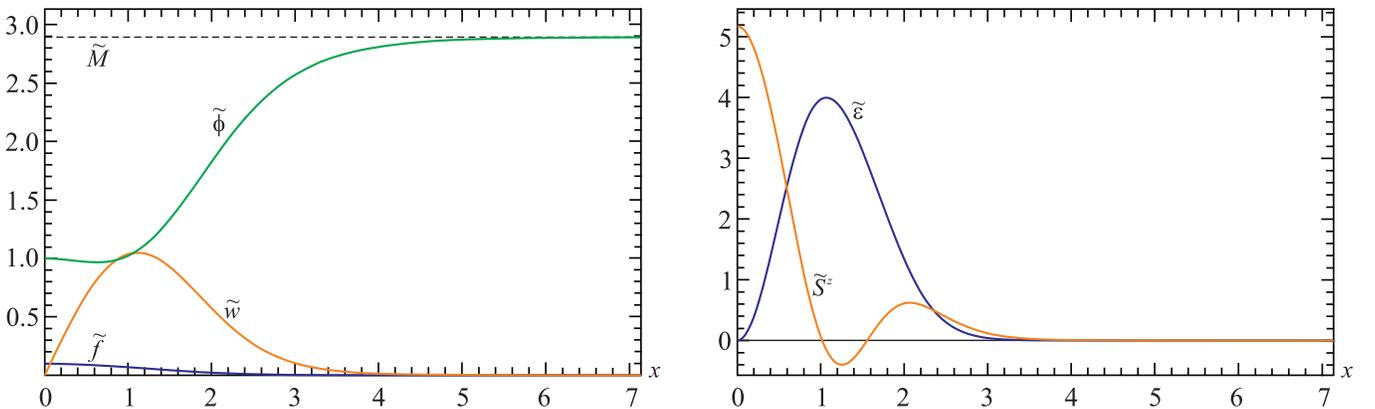}
\caption{The typical profiles of the functions $\tilde f(x), \tilde w(x)$, $\tilde \phi(x)$, the dimensionless energy $\tilde \varepsilon(x)$, and the momentum density $\tilde S^z (x)$
for
$\tilde \lambda =2.0, \tilde\Lambda = 0.1$,
$\tilde\phi_0 = 1.0, \tilde w_1 = 1.5, \tilde f_0 = 0.1$,
$
	\tilde\mu = 1.53212, \tilde\mu_3 = 1.92906, \tilde M = 2.89307$.
}
\label{n_Abelian_f_w_phi_FT_momentum_energy_density}
\end{figure}

The typical profiles of the functions $\tilde f(x), \tilde w(x)$, and $\tilde \phi(x)$
and of the dimensionless energy density
\begin{equation}
\begin{split}
	\tilde \varepsilon\equiv \frac{g^2}{\phi^4_0}  \varepsilon = &
	\frac{\left( \tilde f^\prime \right)^2}{2} +
	\frac{\left( \tilde v^\prime \right)^2}{2} +
	\frac{1}{2} \left(
		\tilde w^\prime + \frac{\tilde w}{x}
	\right)^2 +
	\frac{1}{\tilde \lambda} \frac{\left( \tilde{\phi}^\prime \right)^2}{2} +
	\frac{\tilde w^2}{8} \left( \tilde f^2 + \tilde v^2 \right) -
	\frac{\tilde \mu_1^2}{2} \tilde f^2 -
	\frac{\tilde \mu_2^2}{2} \tilde v^2 -
	\frac{\tilde \mu_3^2}{2} \tilde w^2
\\
	&
	+\frac{{\tilde \phi}^2}{2} \left(
		 \tilde f^2 + \tilde v^2 + \tilde w^2
	\right) +
			\frac{\tilde \Lambda}{4 \tilde \lambda}
			\left( \tilde \phi^2 - \tilde M^2 \right)^2
\end{split}
\label{4_b_90}
\end{equation}
are shown in Fig.~\ref{n_Abelian_f_w_phi_FT_momentum_energy_density}.
Substitution of the components of electric and magnetic fields \eqref{3_b_20} and \eqref{3_b_30} in Eq.~\eqref{3_b_35}
yields the following expression for the Poynting vector:
\begin{equation}
	S^z = \frac{1}{g^2} \left(
		\frac{d f}{d \rho} \frac{d v}{d \rho} + \frac{1}{4}f v w^2
	\right)  =
	\frac{ \phi^4_0}{g^2} \left(
		\tilde f^\prime \tilde v^\prime +
		\frac{1}{4}\tilde f \tilde v \tilde w^2
	\right) =
	\frac{ \phi^4_0}{g^2} \tilde S^z
	=\left(\frac{\phi_0^4/\Lambda^4_{\text{QCD}}}{g'^2}\right)\hbar c\Lambda^4_{\text{QCD}}\tilde S^z ,
\label{3_b_50}
\end{equation}
whose distribution is given in Fig.~\ref{n_Abelian_f_w_phi_FT_momentum_energy_density}.
Note that this expression contains the gradient term
$\tilde f^\prime \tilde v^\prime$ [which is the same as that given in Eq.~\eqref{4_a_60}] and the nonlinear term $\tilde f \tilde v \tilde w^2/4$, which appears because the Proca field is non-Abelian.

\begin{figure}[H]
\centering
	\includegraphics[width=1\linewidth]{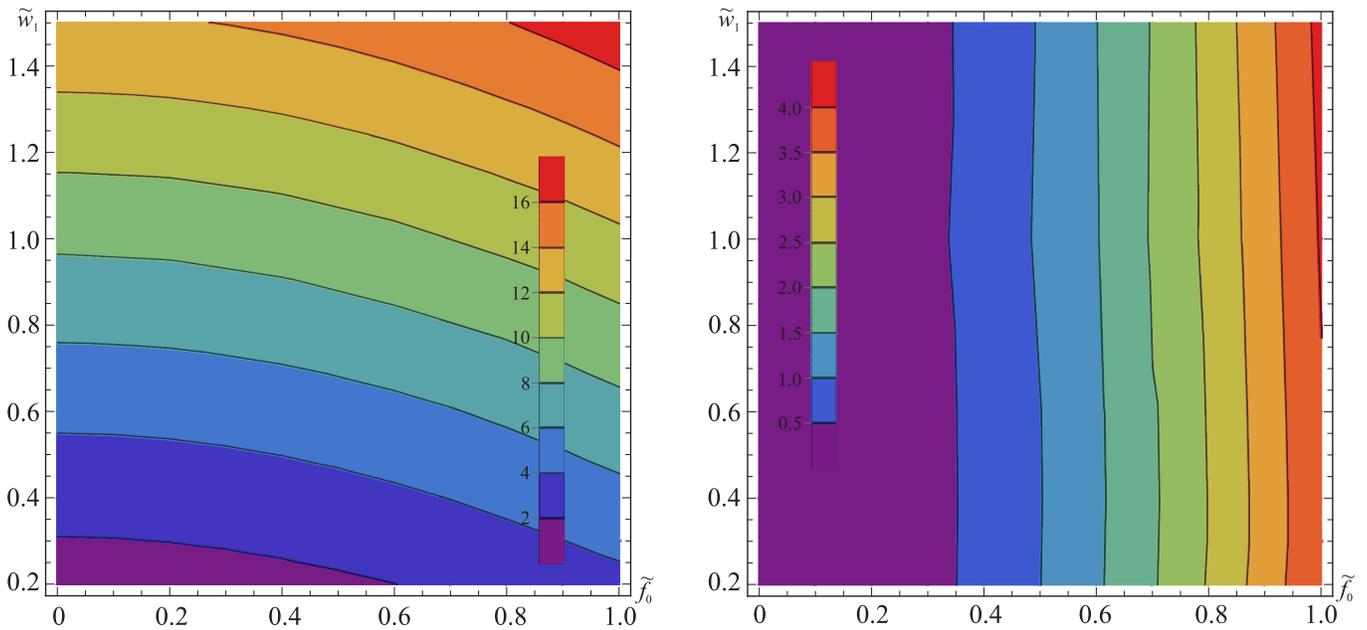}
\caption{The contour profiles of the linear energy density
	$\tilde{\mathcal E}$ (left panel) and of the linear momentum density  $\tilde \Pi^z$  (right panel) on the parameters $\tilde f_0$ and $\tilde w_1$.
}
\label{n_Abelian_linear_energy_momentum}
\end{figure}

Similarly to Sec.~\ref{Abelain_Proca_FT}, we have calculated the integral characteristics of the non-Abelian tube filled with the color electric and
magnetic Proca fields. The results are given in Fig.~\ref{n_Abelian_linear_energy_momentum}.

Thus, we have shown in this subsection that in SU(3) non-Abelian Proca theory there exist tubes filled with stationary, crossed color
electric and magnetic fields, which, by virtue of being crossed, create the energy flux and momentum directed along the tube axis. These solutions are topologically trivial.

The analysis of the results obtained permits us to assume that
\begin{itemize}
\item the linear momentum density  $\Pi^z$ depends weakly on $w_1$, which is the value of the potential
$A^7_\varphi$ on the tube axis;
\item the linear energy density $\mathcal E \rightarrow \text{const}$ for $f_0 \rightarrow 0, w_1 = \text{const}$;
\item the linear momentum density $\Pi^z \rightarrow 0$ for $f_0 \rightarrow 0, w_1 = \text{const}$;
\item $\mathcal E, \Pi^z \rightarrow \infty$ as $w_1 \rightarrow \infty$.
\end{itemize}

\section{Conclusions}
\label{concl}

In the present paper, we continued our investigations begun in Refs.~\cite{Dzhunushaliev:2019sxk,Dzhunushaliev:2020eqa} concerning tube solutions within SU(3) non-Abelian-Proca-Higgs theory. Our purpose was to obtain the integral characteristics of those solutions. We were interested in obtaining the dependencies of the flux of longitudinal electric field, and linear momentum/energy densities on the system parameters.

We have considered two types of tubes. For the tubes of the first type, possessing the flux of  longitudinal color electric field, we have obtained the dependencies of the flux and of the linear energy density
on the value of the coupling constant $\Lambda$ and on the values of the fields at the center of the tube. For the tubes of the second type, possessing the momentum density, we have found the dependencies of the linear  momentum and energy densities on the values of the fields at the center of the tube.

The results obtained can be summarized as follows:
\begin{itemize}
\item For the tube with the flux of color electric Proca field directed along the tube axis, we have studied the dependence of the linear energy density and of the flux of such field on the values
of the components $A^2_t$ and $A^5_z$ on the tube axis.
\item For the tube with the energy flux (and hence with the momentum directed along the tube axis), we have examined the dependence of the linear energy and momentum densities on the values of
the components $A^5_{t,z}$ and $A^7_\varphi$ on the tube axis.
\item The existence of the aforementioned tube solutions depends crucially on the presence of the Higgs scalar field singlet (there are no such solutions without this field).
\item The tube solutions obtained are topologically trivial, in contrast to the Nielsen-Olesen solution~\cite{Nielsen:1973cs} carrying a topological charge.
\item The solution with the color longitudinal electric field demonstrates the dual Meissner effect: the electric field is pushed out by the Higgs scalar field.
\end{itemize}

QCD is SU(3) non-Abelian Yang-Mills theory containing one dimensional constant
$\Lambda_{\text{QCD}}=200~\text{MeV}/ \hbar c  \approx 1~\text{fm}^{-1} \approx 10^{13}\text{cm}^{-1} $.
Here, we have considered SU(3) non-Abelian Proca theory, which differs from QCD by having massive terms.
It is seen from the corresponding formulae that all the integral characteristics calculated here depend on $\phi_0$, whose dimension is
$[\phi_0] = \text{cm}^{-1}$. For this reason, it is interesting to estimate the values of the integral quantities under consideration for
$\phi_0 \approx \Lambda_{\text{QCD}}$. In this case the dimensional coefficients appearing in the expressions
\eqref{3_a_170} and \eqref{3_a_180} are
\begin{equation*}
			\left(
			\frac{\phi_0/\Lambda_{\text{QCD}}}{g'}
		\right)^2\hbar c \Lambda^2_{\text{QCD}} \approx
		200~\text{MeV}/\text{fm}
		\approx 3.2\cdot 10^9 \text{erg}/\text{cm}
		\approx 3 \text{ tones};
\end{equation*}
the latter value is of the same order as the force of interaction between quarks which appears as a consequence of the existence of a flux tube
filled with a SU(3) gauge Yang-Mills field and connecting the quarks.

In conclusion, we would like to point out that in order to understand the nature of confinement in QCD, it is necessary to have flux tubes
filled with a color longitudinal electric field which connect quarks and create a linear potential between quarks ensuring confinement.
Another interesting feature of QCD is that gluon fields give a considerable contribution to the proton spin.
As we demonstrated in Refs.~\cite{Dzhunushaliev:2019sxk, Dzhunushaliev:2020eqa} and in the present paper, in
SU(3) non-Abelian Proca theory, there are two types of tube solutions: (i) with the flux of color electric field (in Proca theory, such tubes connect ``quarks'' and may lead to
confinement in massive Yang-Mills theories); (ii) with the momentum directed along the tube (if such tubes connect ``quarks'' in a Proca proton,
they will contribute to the Proca proton spin). This means that massive non-Abelian Yang-Mills theories
(non-Abelian Proca theories) have some similarities to quantum chromodynamics; this gives good reason to detailed study of such theories.
One more motivation for studying cylindrical solutions  supported by massive vector fields might be the possibility of using them in modelling 
such gravitating configurations like cosmic strings both within Einstein's gravity~\cite{Vilenkin2000} and in various modified gravities~\cite{Nojiri:2017ncd}.

\section*{Acknowledgments}
We gratefully acknowledge
support provided by Grant
in Fundamental Research in Natural Sciences by the Ministry of Education and Science of the Republic of Kazakhstan. We are also grateful to
the Research Group Linkage Programme of the Alexander von Humboldt Foundation for the support of this research.

\end{document}